\documentclass[%
 reprint,
superscriptaddress,
 amsmath,amssymb,
 aps,
 pra,
]{revtex4-1}

\usepackage{graphicx}% Include figure files
\usepackage{dcolumn}% Align table columns on decimal point
\usepackage{bm}% bold math

\newcommand{\ket}[1]{\vert{#1}\rangle}
\begin{document}

\preprint{APS/123-QED}

%\textheight 24cm
%\textwidth 17cm
%\oddsidemargin 0cm
%\evensidemargin -1cm
%\topmargin -1cm
%opening
\title{Quantum dynamics by the constrained adiabatic trajectory method}
\author{A. Leclerc}
\email{Arnaud.Leclerc@utinam.cnrs.fr}
\affiliation{Institut UTINAM, CNRS UMR 6213, Universit\'e de Franche-Comt\'e, Observatoire de Besan\c con,\\
41 bis Avenue de l'Observatoire, BP 1615, 25010 Besan\c con cedex, France}
\author{S. Gu\'erin}
\affiliation{Laboratoire Interdisciplinaire Carnot de Bourgogne \\(CNRS UMR 5209, Universit\'e de Bourgogne),
BP 47870, 21078 Dijon, France}
\author{G. Jolicard}
\affiliation{Institut UTINAM, CNRS UMR 6213, Universit\'e de Franche-Comt\'e, Observatoire de Besan\c con,\\
41 bis Avenue de l'Observatoire, BP 1615, 25010 Besan\c con cedex, France}
\author{J. P. Killingbeck}
\affiliation{Centre for Mathematics, University of Hull, Hull HU6 7RX, UK}

%\date{}

\begin{abstract}
We develop the constrained adiabatic trajectory method (CATM) which allows one to solve the time-dependent Schr\"odinger equation constraining the dynamics to a single Floquet eigenstate, as if it were adiabatic. This constrained Floquet state (CFS) is determined from the Hamiltonian modified by an artificial time-dependent absorbing potential whose forms are derived according to the initial conditions. The main advantage of this technique for practical implementation is that the CFS is easy to determine even for large systems since its corresponding eigenvalue is well isolated from the others through its imaginary part.
The properties and limitations of the CATM are explored through simple examples.
\end{abstract}
\maketitle
\section{Introduction}

Modern developments and applications of quantum mechanics often involve complex chemical and even biological systems driven by laser fields (see for instance \cite{Assion}). Solving (numerically) the time dependent Schrodinger equation (TDSE) for such time dependent systems becomes then very time consuming and sometimes even impossible.
Finding numerical simplifications is an active research. One can for instance mention the multi-configuration time-dependent Hartree (MCTDH) algorithm \cite{Meyer}.

Techniques that lead to an efficient propagation of a time-dependent problem often involve the Floquet theory which allows one to incorporate fast oscillations of the external field (for instance such as the optical oscillations of a laser field) in an enlarged Hilbert space \cite{shirley}. For instance, it permits an adiabatic separation between the fast field oscillation dynamics and the slow time modulation
of the field envelope (adiabatic Floquet theory \cite{reviewguerin}). This Floquet technique can be used alternatively to treat the full time-dependence of the field, which is referred to as the $(t,t')$ approach \cite{peskin}.

Relevant processes are most often expected to be described in a small subspace, often named active space, through effective Hamiltonians. One can mention in particular the time-dependent wave operator theory (TDWOT) as a tool to extract dynamical active spaces \cite{reviewgeorges2}.

A few years ago Jolicard et al. \cite{CATM} have proposed the ``Constrained Adiabatic Trajectory Method'' (CATM) 
for solving the TDSE for a time-dependent potential.
Since we use a quantum mechanical approach the trajectory studied in the
CATM is not a classical one but rather a constrained path followed by the wavefunction as it develops in time in a
composite Hilbert space which we describe below.
We here investigate that method
extending it for an initial condition as a general superposition of states for a small system,
and emphasizing its principal novel feature, the
use of a complex absorbing potential which is itself time-dependent.
The usual approach is to propagate the wavefunction in small time steps, with the Hamiltonian
considered as constant over each step \cite{leforestier,kosloff}. The radically different approach of the CATM
is to limit the time development to only one term in a Floquet expansion of the wavefunction,
achieving this by a careful choice of the varying absorbing potential.
The problem of integrating the TDSE then becomes that of finding one eigenvector of the system's
Floquet Hamiltonian. The method has some affinities with the $(t,t')$ approach \cite{peskin}
but represents a modification and improvement of it. The method finds the wavefunction
at regular grid points throughout the interaction period, the principal requirement being to work
with a sufficient number of points to describe the time-varying Hamiltonian and
to allow the stable use of Fast Fourier transforms.

In brief, the technique requires the wavefunction corresponding to the dynamics to connect to a single Floquet state, referred to as a constrained Floquet state (CFS), through the use of an artificial absorbing potential (or optical potential). The second role of the absorbing potential is to dilate the Floquet spectrum isolating well the eigenvalue corresponding to the CFS from the other ones. In practice one thus needs to find this CFS to determine the dynamics.

In Section II, on the basis of Ref. \cite{CATM}, we summarize the technique CATM with its corresponding Floquet representation, and recall the result when the initial condition is a single eigenstate of the free system. In Section III, we extend the technique to a more general initial state, as a superposition of eigenstates of the free system. This is analyzed for a two-state system.
A forthcoming work will treat the case of systems of higher dimension.
In Section IV, we give an analytic treatment of the effect of the absorbing potential on the Floquet spectrum for a two-state model. The numerical limitations of the method and its accuracy are analyzed in section V.
We study examples with two- and three-level models which illustrate the dual role played by the optical potential in Section VI. Section VII is devoted to the conclusion.

\section{The constrained adiabatic trajectory method}
\subsection{The Floquet representation}

We assume a system of Hamiltonian $H(q,t)$ (where the quantum coordinates have been denoted by $q$) defined in a basis $\{|j\rangle\}$. This Hamiltonian can be usually decomposed as $H(q,t)=H_0(q)+W(q,t)$ featuring a free system $H_0(q)$ subjected to an external time dependent field corresponding to the interaction potential $W(q,t)$. In that case, $\{|j\rangle\}$ correspond to the states of the free system. We assume that the interaction potential $W(q,t)$ acts on a duration $t\in[0,T]$ referred to as the physical duration in the following.
We define an extra time interval $[T,T']$ after the physical interaction time during which (i) we add an artificial time-dependent absorbing (or optical) potential ${\cal{V}}(q,t)$ satisfying
${\cal{V}}(q,0\le t\le T)={\cal{V}}(q,T')=0$, and (ii) we extend continuously the interaction such that $W(q,T')=W(q,0)$. This construction features a periodic Hamiltonian $H(q,T')=H(q,0)$.

We can thus define the corresponding Floquet Hamiltonian (or quasi-energy operator) on the extended Hilbert space (product
of the original Hilbert space, i.e. associated to the free system, by the space of $T'$-periodic functions)~\cite{reviewguerin}:
\begin{equation}
 H_F(q,t)=H_0(q)+W(q,t)+{\cal{V}}(q,t)-i\hbar \frac{\partial}{\partial t}.
\label{hamiltonian}
\end{equation}
We consider the entire duration of the interaction+absorbing potential as a fundamental period $T'$ ($\omega_0 = 2\pi / T'$), contrary
to the traditional Floquet scheme in which $T'$ is associated with the period of an external field (such as the optical period of a laser field).
The Floquet states can be indexed with a double labelling $j,n$ linked to a finite basis representation of the decoupled parts of \eqref{hamiltonian}, i.e. to
the free-system eigenstates
 ($j\leftrightarrow \vert j \rangle$) and to the operator $-i\hbar \partial_t$ (corresponding here to a Fourier basis, $n \leftrightarrow |n\rangle\equiv\vert e^{-i n \omega_0 t} \rangle$). %\cite{reviewgeorges2}.
Thus a complete basis is formed with the eigenstates $\{ \vert \lambda_{j,n} (q,t) \rangle\}$~of~$H_F$~:
\begin{equation}
 H_F \vert  \lambda_{j,n} (q,t) \rangle  = \hbar\omega_{\lambda_{j,n}}  \vert \lambda_{j,n} (q,t) \rangle.
\label{eq2}
\end{equation}
Using the periodicity properties of the Floquet theorem ($\ket{\lambda_{j,n} (q,t)}=\ket{\lambda_{j,0} (q,t)}e^{in\omega_0t}$, $\omega_{\lambda_{j,n}}=\omega_{\lambda_{j,0}}+n\omega_0$),
we can rigorously expand the solution of the time dependent Schr\"odinger equation (TDSE)
with an initial limitation to the first Brillouin zone \cite{shirley},
\begin{eqnarray}
\label{dev}
\vert \Psi (q,t)\rangle &=& \sum_{j}   \langle \lambda_{j,0} (q,0)  \vert \Psi (q,0) \rangle
%\nonumber \\
% & &\times \;
e^{-i \omega_{\lambda_{j,0}} t} \vert \lambda_{j,0} (q,t) \rangle.\quad\
\end{eqnarray}
(Here we consider for simplicity only a bound spectrum, that can feature however imaginary parts;
 the extension to a system with bound and continuous spectrum is in principle direct assuming a discretization of the continua).
Usually, a great number of $\vert \lambda_{j,0} \rangle$ is necessary to reconstruct $\vert \Psi (q,t) \rangle$. An interesting practical application of Eq. \eqref{dev}
is the development of a very reduced number of Floquet vectors, and in
the best case of only one, which is the key idea of the CATM.

The method developed in this paper deals with the case of a single constrained Floquet state (CFS) and labeled $\ell$ in the expansion \eqref{dev}.
In this case, the CFS has to match,
when projected at $t=0$, with the initial boundary conditions required for the wavefunction $\Psi(t=0)$:
\begin{equation}
\langle \lambda_{j,0} (q,0)  \vert \Psi (q,0) \rangle=\delta_{j,\ell},
\end{equation}
i.e.
\begin{equation}
 \vert \Psi(q,t) \rangle = e^{-i \omega_{\lambda} t } \vert \lambda (q,t) \rangle,
\label{wavefunction}
\end{equation}
where we have omitted in the latter the index $\ell$ for simplicity: $\omega_{\lambda}\equiv  \omega_{\lambda_{\ell,0}}$, $\vert \lambda (q,t) \rangle\equiv \vert \lambda_{\ell,0} (q,t) \rangle$.

We will show below that in practice we do not get the exact equality \eqref{wavefunction} but a proportionality through a well defined complex phase.

\subsection{Initial condition as an eigenstate of the free system}

Jolicard et al. \cite{CATM} provided the matching with the initial condition for
an initial state equal to a single state $\vert i \rangle$ of the $\{\vert j \rangle\}$ basis, i.e. $\vert \Psi (q,0) \rangle = \vert i \rangle$.
The connection between the Floquet eigenstate and the required initial state is made thanks to the addition of the absorbing potential ${\cal{V}}$ on the extra interval $[T,T']$.
Below we summarize this scheme and extend it in the next section to any required initial condition for the particular case of a two-state system.

In order to satisfy Eq.(\ref{wavefunction}) (with a proportionality instead of the equality), it is sufficient to have the connection at $t=0$:
\begin{equation}
\vert \lambda (q,0)\rangle \propto \vert i \rangle.
\label{CI1}
\end{equation}
We remark that $\vert \lambda (q,t)\rangle$ is a Floquet vector of the extended Hilbert space, but fixing $t$ to a specific value leads to a component of this vector of dimension of the original Hilbert space.
Eq.(\ref{CI1}) suggests the use of the following form for the absorbing potential:
\begin{equation}
 {\cal{V} }(t) = \sum_{j\neq i} -i V_{\text{opt}}(t) \vert j \rangle \langle j \vert
\label{potentiel}
\end{equation}
with $V_{\text{opt}}(t)$ zero over $[0,T]$ and positive over $[T,T']$.
As shown in \cite{CATM}, provided that
\begin{equation}
\label{cond}
\frac{1}{\hbar} \int_{T}^{T'} V_{\text{opt}}(t) dt \gg |\Im (\omega_{\lambda})| (T'-T)
\end{equation}
we can be sure that all channel except $\vert i \rangle$
are absorbed and that Eq.(\ref{CI1}) is satisfied to a good approximation (as will be tested in section \ref{inspection}).

\section{Extension of the CATM to a general initial condition: The two-state case}
\label{Extension}
If we wish to work with CATM in the case of an initial condition as a state superposition of the free system, i.e.
\begin{equation}
\ket{\Psi(0)} = \sum_{j}c_j \ket{j},
\end{equation}
then simple forms as (\ref{potentiel}) no longer work. (From now on, we do not write explicitly the dependence on the $q$ coordinates.)

We provide below the relevant absorbing potential that should be used for a two-state system of Hamiltonian
 \begin{equation}
H(t)=\hbar\left(
\begin{array}{ll}
\Delta_1(t) & \Omega(t) \\
\Omega^{\ast}(t) & \Delta_2(t)
 \end{array}
\right).
 \end{equation}
%or more conveniently written in the interaction representation
We consider the most general case with diagonal terms that are time dependent (due to Stark shifts of the states for instance) and complex (i.e. including their lifetime).
We assume that the coupling $\Omega(t)$ is in general different from zero only over the physical time interval $[0,T]$. During the extra time interval, the diagonal terms have to be continuously varied such that they recover their initial value in order to guarantee the periodicity: $\Delta_j(T')=\Delta_j(0)$.

We show below that it is possible to treat any initial condition by adding
the following absorbing potential over the supplementary interval $[T,T']$:
 \begin{eqnarray}
\label{potgen}
{\cal{V}}(t) & =& \left(
\begin{array}{ll}
0 & 0 \\
 - \frac{c_2e^{i\int_t^{T'}\Delta_2(t')dt'}}{c_1e^{i\int_t^{T'}\Delta_1(t')dt'}} & 1
 \end{array}
\right) \times \left(-i V_{\text{opt}}(t)\right) \\
\text{ with } & &V_{\text{opt}}(t) > 0 \quad \forall t \in ]T,T'[ \nonumber\\
& &V_{\text{opt}}(t) = 0 \quad \forall t \in [0,T]. \nonumber
\end{eqnarray}
The operator
 \begin{eqnarray}
\Pi(t) & =& \left(
\begin{array}{ll}
0 & 0 \\
 - \frac{c_2e^{i\int_t^{T'}\Delta_2(t')dt'}}{c_1e^{i\int_t^{T'}\Delta_1(t')dt'}} & 1
 \end{array}
\right)
\end{eqnarray}
involved in this definition \eqref{potgen} is a non-orthogonal (i.e. non self-adjoint) projector, i.e. $\Pi^2=\Pi$, whose kernel is the initial state up to a phase correction:
 \begin{equation}
\Pi(t)\left(
\begin{array}{c}
c_1 e^{i\int_t^{T'}\Delta_1(t')dt'}\\
c_2e^{i\int_t^{T'}\Delta_2(t')dt'}
 \end{array}
\right)=0.
 \end{equation}
For this case it is indeed possible to obtain the analytical asymptotic form of the Floquet eigenvector
over the extra interval $[T,T']$, where the Hamiltonian contains just the free system Hamiltonian and
the absorbing potential. With the above definition and writing Floquet components
$\langle j \vert \lambda (t) \rangle = \lambda_j (t)$, one must solve on $[T,T']$
the following system :
\begin{subequations}
\begin{eqnarray}
%\left \lbrace
%\begin{array}{ll}
\frac{\partial \lambda_1 (t)}{\partial t} &=& i(\omega_{\lambda}-\Delta_1(t)) \lambda_1 (t) \\
\frac{\partial \lambda_2 (t)}{\partial t} &=&
\frac{V_{\text{opt}}(t)}{\hbar} \frac{c_2e^{i\int_t^{T'}\Delta_2(t')dt'}}{c_1e^{i\int_t^{T'}\Delta_1(t')dt'}} \lambda_1 (t) \nonumber\\
&&- \left(
\frac{V_{\text{opt}}(t)}{\hbar} -i\left(\omega_{\lambda}-\Delta_2(t)\right)\right) \lambda_2 (t)\qquad
%\end{array}
%\right.
\end{eqnarray}
\end{subequations}
The first component follows an exponential law:
$ \lambda_1 (t) = \lambda_1(T) e^{i(\omega_{\lambda}(t-T)-\int_T^t\Delta_1(t')dt')}$.
This function can be introduced in the second equation and making use of
the identity $\int_T^t V_{\text{opt}}(t') e^{\frac{1}{\hbar} \int_T^{t'} V_{\text{opt}}(t'') dt''}
dt'=\hbar \left( e^{\frac{1}{\hbar} \int_T^t V_{\text{opt}} (t') dt'}-1\right)$, we find
\begin{align}
 \lambda_2(t)&=\lambda_2(T) e^{i (\omega_{\lambda}(t-T)-\int_T^t\Delta_2(t')dt')} e^{-\frac{1}{\hbar}\int_T^t V_{\text{opt}}(t')dt'} \nonumber\\
&+\frac{c_2}{c_1} \lambda_1(T) e^{i\omega_{\lambda}(t-T)}
 e^{i\int_{T'}^T\Delta_1(t')dt'}e^{i\int_t^{T'}\Delta_2(t')dt'}\nonumber\\
 &\times(1-e^{-\frac{1}{\hbar}\int_T^t V_{\text{opt}}(t')dt'}).
\label{expdec}
\end{align}
Taking into account the $\lambda$ periodicity $\lambda_j(T')\equiv \lambda_j(0)$, we obtain
\begin{align}
\label{ratio21}
\frac{\lambda_2(0)}{\lambda_1(0)}&=\frac{\lambda_2(T)}{\lambda_1(T)} e^{-\frac{1}{\hbar}\int_T^{T'} V_{\text{opt}}(t)dt} e^{i\int_T^{T'}(\Delta_1(t)-\Delta_2(t))dt}\nonumber\\
&+\frac{c_2}{c_1} (1-e^{-\frac{1}{\hbar}\int_T^{T'} V_{\text{opt}}(t)dt}),
\end{align}
which, in the limits
\begin{subequations}
\label{cond_}
\begin{align}
\label{cond_1} \frac{1}{\hbar}\int_{T}^{T'} V_{\text{opt}}(t) dt &\gg 1,\\
\label{cond_2} \frac{1}{\hbar}\int_{T}^{T'} V_{\text{opt}}(t) dt &\gg \int_T^{T'}[\Im(\Delta_2(t))-\Im(\Delta_1(t))]dt
\end{align}
\end{subequations}
and for $\lambda_2(T)$ and $\lambda_1(T)$ of the same order, leads to
\begin{equation}
\frac{\lambda_2(0)}{\lambda_1(0)}\leadsto\frac{c_2}{c_1}.
\end{equation}
We remark that, denoting the state-vector of the original TDSE $\ket{\Psi(t)}\equiv\left(\begin{array}{ll}
a_1(t)\\
 a_2(t)
 \end{array}\right)$, the connection to a single Floquet vector
\eqref{wavefunction} leads to $\lambda_2(T)/\lambda_1(T)=a_2(T)/a_1(T)$, i.e. to the ratio of the amplitude at the end of the process. If this ratio becomes very large, which corresponds to the specific case of an efficient population transfer to state 2, the condition \eqref{cond_2} is not sufficient. It should be replaced in general by the condition:
\begin{align}
\label{cond_2_} \frac{1}{\hbar}\int_{T}^{T'} V_{\text{opt}}(t) dt &\gg \int_T^{T'}[\Im(\Delta_2(t))-\Im(\Delta_1(t))]dt\nonumber\\
&+\log(a_2(T))-\log(a_1(T)).
\end{align}
 This is discussed in more detail in Section V.B.

For an initial condition as a single state of the free system, ie. $c_2=0$, $c_1=1$, one recovers $\lambda_2(0)\ll\lambda_1(0)$ \cite{CATM}.
In this case, we must note that conditions (\ref{cond_}) are less restrictive than condition (\ref{cond}).
This is due to the fact that conditions (\ref{cond_}) are obtained constraining a ratio of two components, whereas in \cite{CATM} we wished to absorb the components, with an error lower than the computer accuracy.
If the conditions \eqref{cond_} are satisfied, then we can force any eigenstate $\ket{\lambda}$
to obey the final condition
\begin{subequations}
\begin{eqnarray}
%\left \lbrace
%\begin{array}{ll}
\lambda_1(0) &=& \lambda_1(T)e^{i(\omega_{\lambda}(T'-T)-\int_T^{T'}\Delta_1(t)dt)}  \\
\lambda_2(0) & \leadsto & \lambda_1(T) e^{i(\omega_{\lambda}(T'-T)-\int_T^{T'}\Delta_1(t)dt)} \frac{c_2}{c_1}
%\end{array} \right.
\end{eqnarray}
\end{subequations}

Thus, apart from a global constant $\lambda_1(T)$ which results from the diagonalization procedure, an exponentially decreasing term and a global phase,
we obtain
\begin{equation}
\ket{\lambda (t=0)} \propto \ket{\psi (t=0)}.
\end{equation}
This approximate proportionality is sufficient to impose
the required initial connection to the Floquet eigenvector \eqref{CI1}. This will be illustrated
by an example given in section \ref{inspection}.

\section{Isolating one eigenvalue in the Floquet spectrum \label{eigenvalues}}

The second role of the absorbing potential is to dilate the Floquet spectrum and so isolate the ``connected'' eigenvalue $\hbar\omega_{\lambda}$ (i.e. the one associated to the eigenvector $\vert \lambda(t) \rangle$ connected to the initial condition) from the other eigenvalue (denoted $\hbar\omega_{\lambda'}$ associated to $\vert \lambda' (t) \rangle$).
We consider for simplicity the initial condition as a single bound state $\ket{1}$ of $H_0$. The absorbing potential takes the form set out in Eq. (\ref{potentiel}).

We start connecting the solution $\vert \Psi (q,t)\rangle$ to the Floquet vector. This is achieved by solving the stationary problem (in the first Brillouin zone):
\begin{subequations}
\label{0Tprime}
\begin{eqnarray}
\label{0T}
&&\text{for $t\in[0,T]$}:\nonumber\\
&&\left[- i \frac{\partial}{\partial t}+\left(
\begin{array}{ll}
\Delta_1(t) & \Omega(t) \\
\Omega^{\ast}(t) & \Delta_2(t)
 \end{array}
\right)\right]\ket{\lambda(t)}=\omega_{\lambda}\ket{\lambda(t)},\\
&&\text{for $t\in[T,T']$}:\nonumber\\
\label{TTprime}& &\left[- i \frac{\partial}{\partial t}+\left(
\begin{array}{ll}
\Delta_1(t) & 0 \\
0 & \Delta_2(t)-\frac{i}{\hbar}V_{\text{opt}}(t)
 \end{array}
\right)\right]\ket{\lambda(t)}=\omega_{\lambda}\ket{\lambda(t)}.\nonumber\\
\end{eqnarray}
\end{subequations}
In the region $t\in[T,T']$, we obtain from \eqref{TTprime} (see the preceding section):
\begin{subequations}
\begin{eqnarray}
\lambda_1 (t) &=& \lambda_1(T) e^{i(\omega_{\lambda}(t-T)-\int_T^t\Delta_1(t')dt')},\\
\lambda_2 (t) &=& \lambda_2(T) e^{i(\omega_{\lambda}(t-T)-\int_T^t\Delta_2(t')dt')}
e^{-\frac{1}{\hbar} \int_T^t V_{\text{opt}} (t') dt'}.\qquad
\end{eqnarray}
\end{subequations}

\subsection{Decoupled channels}

The situation is the easiest to follow in the
elementary case in which the channels are not coupled ($\Omega(t)=0$) and with constant diagonal terms $\Delta_i$.
Thus we make the instant $T$ coincide with $t=0$,  to study the influence of the optical potential alone
on the interval $[T=0,T']$ without any physical coupling terms.
In this particular case, with $T=0$ and $t=T'$ the previous system becomes:
\begin{subequations}
\begin{eqnarray}
\lambda_1 (T') &=& \lambda_1(0) e^{i(\omega_{\lambda}-\Delta_1)T'},\\
\lambda_2 (T') &=& \lambda_2(0) e^{i(\omega_{\lambda}-\Delta_2)T'}
e^{-\frac{1}{\hbar} \int_0^{T'} V_{\text{opt}} (t') dt'}.\qquad
\end{eqnarray}
\end{subequations}
The same equations can be written for the other eigenstate $\ket{\lambda'}$.
Floquet eigenvectors must be periodic, i.e. $\lambda_i(T') = \lambda_i(0)$. Thus each Floquet eigenvalue
must satisfy simultaneously two conditions:
\begin{subequations}
\begin{eqnarray}
1 &=&  e^{ i (\omega_{\lambda} - \Delta_1)T'} \quad \text{ if} \; \lambda_1(0) \neq 0  \\
1 &=&  e^{ i ((\omega_{\lambda} - \Delta_2)T' + \frac{i}{\hbar} \int_0^{T'} V_{\text{opt}}(t) dt)} \quad \text{if} \;\lambda_2(0)\neq 0
\end{eqnarray}
\end{subequations}
The only solution is to have only one non-zero components for each eigenvector:
\begin{subequations}
\begin{eqnarray}
\lambda_1(0)&\neq&0 \quad \text{and}\quad \lambda_2(0)=0 \quad \text{i.e. } \omega_{\lambda}  =  \Delta_1 \\
\lambda'_1(0)&=&0 \quad \text{and}\quad \lambda'_2(0)\neq0 \nonumber \\
 \text{i.e. } \quad &\omega_{\lambda'}&  =  \Delta_2 - \frac{i}{\hbar T'} \int_0^{T'} V_{\text{opt}}(t)dt
\end{eqnarray}
\end{subequations}
The terms $\frac{2k\pi}{T'}$ are not mentioned because we work in a given Brillouin zone.
In this simpliest case, the extension to a $N-$dimension system is straightforward: all the eigenvalues connected to absorbed channels possess an imaginary term proportional to $\frac{1}{T'}\int_0^{T'} V_{\text{opt}}(t)dt$.
Thus we expect to obtain a dispersion of the eigenvalues in the complex plane which will leave the other eigenvalues
distant from the ``connected`` eigenvalue $\omega_{\lambda}$.
%The case evoked above is evidently too simple,
%nevertheless it might be relevant to a situation in which the absorbing potential is chosen to be very large
%compared to other terms in the Hamiltonian, so that all other coupling terms can be regarded as perturbations.

\subsection{General case}

In the present case of a 2-level coupled system described by Eq.(\ref{0Tprime}), it is possible to go further in the analytical description.
In the region $t\in[0,T]$, one can rewrite \eqref{0T} as
\begin{equation}
\label{0T_}
\left[- i \frac{\partial}{\partial t}+\left(
\begin{array}{ll}
\Delta_1(t) & \Omega(t) \\
\Omega^{\ast}(t) & \Delta_2(t)
 \end{array}
\right)\right]\ket{\lambda(t)}e^{-i\omega_{\lambda}t}=0,\\
\end{equation}
that is as the same form of the original TDSE of solution $\ket{\Psi(t)}\equiv\left(\begin{array}{ll}
a_1(t)\\
 a_2(t)
 \end{array}\right)$.
We connect the two solutions invoking the initial conditions $a_1(0)=1$, $a_2(0)=0$, and $ \lambda_1 (0) = \lambda_1(T) e^{i(\omega_{\lambda}(T'-T)-\int_T^{T'}\Delta_1(t)dt)}$, $\lambda_2(0)\simeq 0$ (from the preceding section):
\begin{equation}
\left(\begin{array}{ll}
a_1(t)\\
 a_2(t)
 \end{array}\right)\lambda_1(T)e^{i(\omega_{\lambda}(T'-T)-\int_T^{T'}\Delta_1(t')dt')}
 =\left(\begin{array}{ll}
\lambda_1(t)\\
 \lambda_2(t)
 \end{array}\right)e^{-i\omega_{\lambda}t}.
 \end{equation}
The latter equation is just the proof of the Floquet theorem for our specific two-state problem. Considering the final physical time $t=T$, we get
\begin{equation}
a_1(T)=e^{i\int_T^{T'}\Delta_1(t)dt}e^{-i\omega_{\lambda}T'},
\label{a_1T}
\end{equation}
that is we connect the imaginary part of the eigenvalue $\omega_{\lambda}$ to the final probability amplitude:
\begin{equation}
\label{Elambda}
\Im{(\omega_{\lambda})}=\frac{1}{T'}\int_T^{T'}\Im{(\Delta_1(t))}dt+\frac{1}{T'}\log(|a_1(T)|).
 \end{equation}
To get the counterpart relation for the other (``non-connected'') eigenvalue $\omega_{\lambda'}$, we reformulate the complete calculation with the adjoint of $H_F(t)$ (using $\partial_t^{\dagger}=-\partial_t$) :
\begin{equation}
 H_F^{\dagger}(t)=H_0^{\dagger}+W^{\dagger}(t)+{\cal{V}}^{\dagger}(t)-i\hbar \frac{\partial}{\partial t}.
\label{hamiltonian_adj}
\end{equation}
of eigenstates $\{ \vert \widetilde\lambda_{j,n} (t) \rangle\}$
\begin{equation}
 H_F^{\dagger} \vert  \widetilde\lambda_{j,n} (t) \rangle  = \hbar\omega_{\lambda_{j,n}}^{\ast}  \vert \widetilde\lambda_{j,n} (t) \rangle,
\end{equation}
where $(.)^{\ast}$ denotes the complex conjugate. For real energies of $H_0$ and real elements in $W(t)$, this latter equation corresponds to the same original problem as before but
with the use of an exponentially diverging potential ${\cal{V}}^{\dagger}(t)$.
%We remark that the vectors $\langle \widetilde\lambda_{j,n} (t) |$ are
%often referred to as left eigenvectors as opposed to the right eigenvectors %$\vert \lambda_{j,n} (t) \rangle$.
%In that case $\omega_{\lambda'}^{\ast}$ take the preceding role of
%$\omega_{\lambda}$, that is it becomes the connected eigenvalue.
We have then for the components of $\vert \widetilde\lambda' (t) \rangle$ (denoted as the eigenvector associated to the eigenvalue $\hbar\omega_{\lambda'}^{\ast}$,
 $\vert \widetilde\lambda' (t) \rangle$ is different from $\vert \lambda' (t) \rangle$ in general):
\begin{subequations}
\begin{eqnarray}
\widetilde\lambda'_1 (t) &=& \widetilde\lambda'_1(T) e^{i(\omega_{\lambda'}^{\ast}(t-T)-\int_T^t\Delta_1^{\ast}(t')dt')},\\
\widetilde\lambda'_2 (t) &=& \widetilde\lambda'_2(T) e^{i(\omega_{\lambda'}^{\ast}(t-T)-\int_T^t\Delta_2^{\ast}(t')dt')}
e^{\frac{1}{\hbar} \int_T^t V_{\text{opt}} (t') dt'},\qquad
\end{eqnarray}
\end{subequations}
which leads in the limits \eqref{cond_} to
\begin{subequations}
\begin{eqnarray}
\widetilde\lambda'_1 (0)& \ll &\widetilde\lambda'_2 (0),\\
\widetilde\lambda'_2 (0) &=& \widetilde\lambda'_2(T) e^{i(\omega_{\lambda'}^{\ast}(T'-T)-\int_T^{T'}\Delta_2^{\ast}(t)dt)} \nonumber\\
&&e^{+\frac{1}{\hbar} \int_T^{T'} V_{\text{opt}} (t) dt}.
\end{eqnarray}
\end{subequations}
It corresponds to the Schr\"odinger equation
\begin{equation}
\label{0Tp}
\left[- i \frac{\partial}{\partial t}+\left(
\begin{array}{ll}
\Delta_1^{\ast}(t) & \Omega^{\ast}(t) \\
\Omega(t) & \Delta_2^{\ast}(t)
 \end{array}
\right)\right]\left(\begin{array}{c}a_1'(t)\\a_2'(t)\end{array}\right)=0\\
\end{equation}
with the initial condition $a_1'(0)=0,a_2'(0)=1$ for which we get
\begin{equation}
a_2'(T)=e^{i\int_T^{T'}\Delta_2^{\ast}(t)dt}e^{-i\omega_{\lambda'}^{\ast}T'}
e^{-\frac{1}{\hbar} \int_T^{T'} V_{\text{opt}} (t) dt}.
 \end{equation}
One can connect it to $a_1(T)$ as described in appendix \ref{appA} which induces
\begin{align}
\label{a_1T_}
a_1(T) =e^{-i\int_0^T[\Delta_1(t)+\Delta_2(t)]dt} e^{-i\int_T^{T'}\Delta_2(t)dt} \nonumber \\
e^{+i\omega_{\lambda'}T'}
e^{-\frac{1}{\hbar} \int_T^{T'} V_{\text{opt}} (t) dt}.
\end{align}
Identifying \eqref{a_1T} and \eqref{a_1T_} leads to
\begin{align}
\Im{(\omega_{\lambda'})}&=\frac{1}{T'}\left(\int_0^{T'}\Im{(\Delta_2(t'))}dt'+\int_0^T \Im(\Delta_1(t'))dt'\right)
\nonumber\\
&-\frac{1}{T'}\log(|a_1(T)|) -\frac{1}{\hbar T'} \int_T^{T'} V_{\text{opt}} (t') dt',
\end{align}
which gives a relation between the imaginary parts of the two eigenvalues:
\begin{align}
\label{Elambdap}
\Im{(\omega_{\lambda'})}&=-\frac{1}{\hbar T'} \int_T^{T'} V_{\text{opt}} (t') dt'
-\Im{(\omega_{\lambda})}\nonumber\\
&+ \frac{1}{T'}\int_0^{T'} \Im(\Delta_1(t')+\Delta_2(t'))dt'.
 \end{align}
This central relation shows that the connected eigenvalue will be in general well isolated from the other one for a large enough area of the absorbing potential. More precisely we have $-\Im{(\omega_{\lambda'})}\gg -\Im{(\omega_{\lambda})}$ when
 \begin{align}
 \label{cond_isol}
&\frac{1}{\hbar T'} \int_T^{T'} V_{\text{opt}} (t') dt'\gg
-2\Im{(\omega_{\lambda})}\nonumber\\
&\qquad\qquad+\frac{1}{T'}\int_0^{T'} \Im(\Delta_1(t')+\Delta_2(t'))dt'.%[\Im{(\Delta_2(T))}+\Im{(\Delta_1(T))}]\frac{T'-T}{T'}
 \end{align}
%that is, when [we also use \eqref{cond_}]
% \begin{equation}
% \label{cond_isol_}
%\frac{1}{\hbar} \int_T^{T'} V_{\text{opt}} (t') dt'\gg
%-2\log(|a_1(T)|))
% \end{equation}
%Unfortunately this important result can't be easily generalized to three- or N-dimension systems because we use a kind of symmetry between the ``connected'' and the ``non-connected'' eigenvalue which disappear when there are more than two eigenvalues in each Brillouin zone.
%However, even without general N-dimension results about the isolation of the ``connected'' eigenvalue,
This feature will be useful in numerical calculations; in particular it will improve the rate of convergence of
the wave operator method \cite{reviewgeorges2} when applied to the location of the thus isolated connected eigenvalue.

However the separation between the imaginary parts of eigenvalues can be not so efficient in practice for specific cases of good population transfer. This is analyzed in the following section.

\section{Numerical limitations and accuracy \label{limitations}}
In this section we study the numerical limitations of the method, restricting the discussion to the situation $c_1(0)=1,c_2(0)=0$.
We consider for simplicity the situation $\Im{(\Delta_2(t))}=\Im{(\Delta_1(t))}=0$.
\subsection{General cases}
The accuracy of the method can be roughly estimated from the imperfect initial connection with the eigenvector $\ket{\lambda}$, that is from the small quantity $\lambda_2(0)$. In general, when $\lambda_2(T)$ and $\lambda_1(T)$ are of the same order, we obtain for the error in the final amplitude from \eqref{ratio21}:
\begin{equation}
\label{acc2}
\left|a_1(T)-a_1^{\text{(CATM)}}(T)\right|\propto e^{-\frac{1}{\hbar} \int_T^{T'} V_{\text{opt}} (t') dt'},
\end{equation}
where $a_1^{\text{(CATM)}}(T)$ is the probability amplitude of state 1 at the end of the physical process obtained from the CATM method.
This is shown to give a correct estimation of the accuracy of the method when it is tested numerically (see next section).

We remark that this estimation does not obviously take into account the grid size effect. This is studied numerically in the next section.

\subsection{Case of good population transfer}
The estimation \eqref{acc2} is not valid when the population transfer at the end of the process is efficient: $|a_1(T)|\to0$, since, in Eq. \eqref{ratio21}, we have then $|\lambda_2(T)/\lambda_1(T)|\gg1$.
The area of the optical potential should be large enough to satisfy the connectivity to a unique Floquet eigenvector: $\lambda_2(0)/\lambda_1(0)\leadsto0$, that is, from \eqref{cond_2_}
\begin{equation}
\label{cond_2__}
\frac{1}{\hbar}\int_{T}^{T'} V_{\text{opt}}(t) dt \gg -\log(a_1(T)).
\end{equation}
One limiting case is when there is no separation between the imaginary parts of the eigenvalues:
\begin{equation}
\label{equalIm}
\Im{(\omega_{\lambda'})}=\Im{(\omega_{\lambda})},
 \end{equation}
leading to
 \begin{align}
&\frac{1}{\hbar}\int_T^{T'} V_{\text{opt}} (t') dt'=
-2\log(|a_1^{\text{(CATM)}}(T)|)).
 \end{align}
This equation shows that, in this case of equal quasienergies, the inequality \eqref{cond_2__} is satisfied with only a factor 2. More precisely, we have
\begin{equation}
\frac{\lambda_2(0)}{\lambda_1(0)}\approx e^{-\frac{1}{2\hbar}\int_{T}^{T'} V_{\text{opt}}(t) dt}.
\end{equation}
Thus, one can still satisfy $\lambda_2(0)/\lambda_1(0)\leadsto0$ to get the connection to a unique Floquet eigenvector to a good accuracy by imposing
\begin{equation}
\label{cond_2___}
\frac{1}{2\hbar}\int_{T}^{T'} V_{\text{opt}}(t) dt \gg 1.
\end{equation}
This condition \eqref{cond_2___}, a bit more restrictive than \eqref{cond_1} is thus sufficient to obtain a quite good relative accuracy of the solution in the case of good population transfer, even if in that case the imaginary parts of the Floquet eigenvalues are close together.

We can use this limiting case \eqref{equalIm} to estimate the absolute accuracy of the method.
Assuming $\Im{(\omega_{\lambda'})}\le\Im{(\omega_{\lambda})}$, we get
\begin{equation}
\label{acc}
|a_1^{\text{(CATM)}}(T)|\ge e^{-\frac{1}{2\hbar} \int_T^{T'} V_{\text{opt}} (t') dt'},
 \end{equation}
that is we cannot obtain numerically a population $|a_1^{\text{(CATM)}}(T)|^2$ of state 1 at the end of the physical process smaller than $e^{-\frac{1}{\hbar} \int_T^{T'} V_{\text{opt}} (t') dt'}$,
which gives thus a numerical limitation of the depopulation of the initial state.

\section{Numerical investigation \label{inspection}}

The method is investigated numerically in this section through the examples of two- and three- state systems driven by a time-dependent field. They can correspond for instance to atoms submitted to resonant laser pulses in the rotating wave approximation (RWA) \cite{shore,guerin}.

\subsection{Some results for selected examples \label{exemples}}

The first example is a two-state system $\{\ket{1},\ket{2}\}$ which is subjected to a pulsed coupling of frequency little detuned with the transition frequency. The detuning is denoted $\Delta$ and $\Omega$ is the coupling (Rabi frequency). In the dressed state picture of the RWA the Hamiltonian is (in units such that $\hbar=1$)
\begin{equation}
H = \left(
\begin{array}{ll}
0 & \Omega  \\
\Omega  & \Delta
\end{array} \right)
= \left(
\begin{array}{ll}
0 & \Omega_0 \sin ^2 \left( \frac{\pi t}{T} \right) \\
\Omega_0 \sin ^2 \left( \frac{\pi t}{T} \right) & \Delta_0 \cos \left( \frac{\pi t}{T} +\phi_0 \right)
\end{array} \right)
\end{equation}
We will consider as initial condition (i) $\ket{\Psi (t=0)} = \ket{1}$,
from which we expect a final quasi-inversion of population for large enough $\Omega_0T$ and $\Delta_0T$ (adiabatic passage, see \cite{shore,guerin}), and (ii) the more complicated situation $\ket{\Psi (0)} = c_1 \ket{1} + c_2 \ket{2}$.

The second example is that of a 3 level system $\{\ket{1} , \ket{2},\ket{3}\}$ driven by two near-resonant laser fields with Rabi frequencies
$\Omega_p$ and $\Omega_s$, tuned to the transitions $1\leftrightarrow 2$ and $2\leftrightarrow 3$ respectively.
We allow a detuning $\Delta$ between the transition frequency $1\rightarrow2$ and the laser frequency and assume a two-photon resonance. The initial state is $\ket{1}$.
Here the RWA Hamiltonian takes the form :
\begin{equation}
H = \left(
\begin{array}{lll}
0 & \Omega_p & 0 \\
\Omega_p & \Delta  & \Omega_s \\
0 & \Omega_s & 0
\end{array} \right)
\end{equation}
We study two situations, on one hand the intuitive case: we first turn on the coupling between levels $1$ and $2$, then
between levels $2$ and $3$,
\begin{eqnarray}
\Omega_p & =& \Omega_0 \sin ^2 \left( \frac{\pi t}{T_1} \right) \quad \forall t \in [0,T_1] \quad (0 \text{ elsewhere}) \nonumber\\
\Omega_s &=& \Omega_0 \sin ^2 \left( \frac{\pi t-T_1/2}{T_1} \right) \quad \forall t \in \left[\frac{1}{2}T_1,\frac{3}{2}T_1 \right] \nonumber\\
\Delta&=&\Delta_0
\end{eqnarray}
With $\Omega_0 T_1= 20$ and $\Delta_0 T_1=0$,
we expect to observe oscillations without complete population exchange to state $\ket{3}$.
With $\Delta_0 T_1=20$ a partial transfer to $\ket{3}$ occurs with less oscillations.

On the other hand the STIRAP case (Stimulated Raman Adiabatic Passage) is exactly the inverse of the first configuration \cite{stirap}:
\begin{eqnarray}
\Omega_p & =& \Omega_0 \sin ^2\left( \frac{\pi t-T_1/2}{T_1} \right) \quad \forall t \in \left[\frac{1}{2}T_1,\frac{3}{2}T_1\right] \nonumber \\
\Omega_s &=& \Omega_0 \sin ^2  \left( \frac{\pi t}{T_1} \right) \quad \forall t \in [0,T_1] \nonumber\\
\Delta & = &0
\end{eqnarray}
With $\Omega_0 T_1= 20$ and $\Delta_0 T_1=0$, STIRAP allows a large transfer of the population to $\ket{3}$.

The total physical time interval $T$ here is $3/2$ times the period $T_1$ of the sine function $[0,T]=[0,3/2T_1]$;
the additional time interval will begin at $3/2T_1$ for a duration of $T_1$.

In the subsequent discussion we use the  labels (i) and (ii) for the 2 level system with initial state $\ket{1}$
and the superposition of states, respectively. The label (iii) and (iv) refer to the 3 level system in the ``intuitive'' or STIRAP situations, respectively.

\subsection{Calculating with CATM}

From a technical point of view the calculation involves the five following steps:
\begin{itemize}
\item Construction of the matrix representation of the Floquet Hamiltonian (some details are given in appendix \ref{appB})
\item Diagonalization of the Floquet matrix
\item Selection of $N$ Floquet eigenstates belonging to the first Brillouin zone (for a problem with $N$ levels)
\item Detection of the appropriate ``connected'' Floquet eigenstate, i.e. corresponding to the smallest imaginary part of the eigenvalue in absolute value as a criterion
\item Production of the wavefunction via Eq. (\ref{wavefunction})
\end{itemize}
In principle only one vector computation is needed. For our small-scale examples we can easily use direct complete diagonalisation. However, for larger systems the time-dependent wave operator can be used to find the required eigenstate of the corresponding large matrix.

\subsection{A comparison with direct integration}

We analyse the results obtained with the Floquet eigenstate which possesses the smallest value
of $\vert \Im (\omega_{\lambda}) \vert$, as predicted by the theory. Next we calculate the populations
\begin{equation}
p_n(t)=\vert \langle n \ket{\Psi (t)} \vert ^2
\end{equation}
 and the relative phases
\begin{equation}
\beta_n(t)=\arg \left( \langle n \ket{\Psi (t)} \right)
\end{equation}
for all the previously presented situations.
We compare the CATM results for the population and phase with those of a direct integration
using the propagation equation
\begin{equation}
\label{Direct}
\ket{ \Psi (t+\Delta t)} = e^{ -i \hbar^{-1} H \left(t+\frac{\Delta t}{2}\right) \Delta t} \ket{ \Psi (t)}
\end{equation}
with $\Delta t$ a sufficiently small time-step.
For the CATM calculation, the size of the Fourier basis set was $N=256$ which is ample for both stable computation and
graphical representation.

\subsubsection{Two-state model}

\begin{figure}[!ht]
 \includegraphics[width=\linewidth]{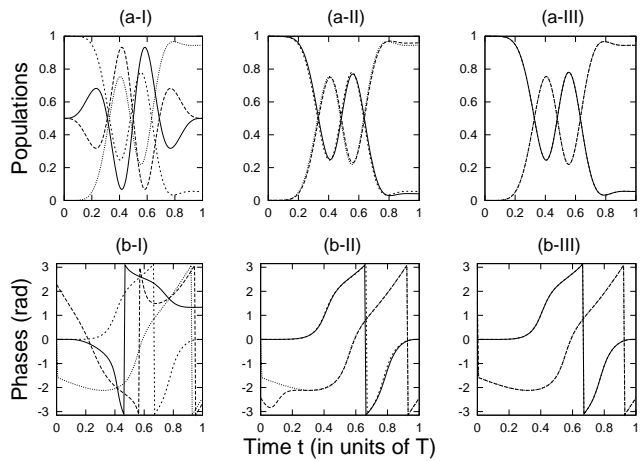}
\caption{Evolution of populations (a) and phases (b) for the 2-level system (i) with the initial state $\ket{1}$, and $\Omega_0 T = 10$ and $\Delta_0 T= 10$.
Exact [i.e. numerical with the direct integration, cf. \eqref{Direct}] results ($p_1$ and $\beta_1$: short dashes, $p_2$ and $\beta_2$: dots)
and CATM results ($p_1$ and $\beta_1$: solid line, $p_2$ and $\beta_2$: long dashes)
for various amplitudes $V_0$ of the time-dependent absorbing potential: (I) $V_0 T=0$ (II) $V_0 T=10$ (III) $V_0 T=40$.}
\label{fig1}
\end{figure}

For the 2 level system (i) the results are shown on Fig. \ref{fig1}.
In frames (a-I) and (b-I), it is evident that without the absorbing potential the use of a single Floquet state is not sufficient. On frames (a-II) and (b-II) we can
observe the effects of the absorbing potential.
The initial populations approach $p_1(0)=1$ and $p_2(0)=0$, showing however a small difference
of a few percent from the reference calculation results. Phases begin to agree with those of the reference calculation
but the difference remains important, especially at the beginning.
For the last case (a-III and b-III), one cannot detect any difference between
the CATM and the reference results at the scale of the figure.

\begin{figure}[!ht]
 \includegraphics[width=\linewidth]{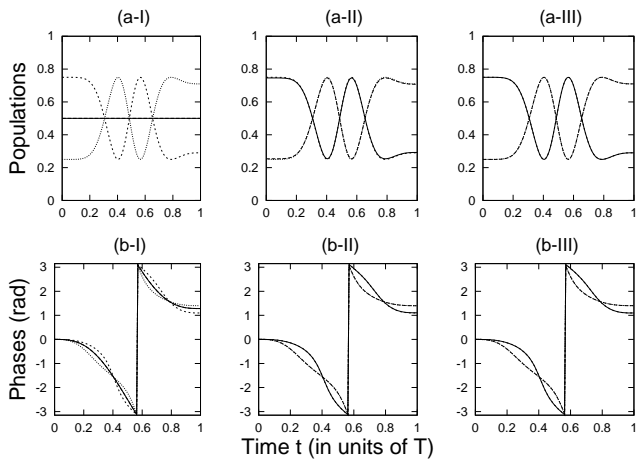}
\caption{Same as Fig. \ref{fig1}, but for the 2-level system (ii), and $\Omega T = 10$, $\Delta_0 T =0$ with the initial state $\sqrt{0.75}\,\ket{1}+\sqrt{0.25}\,\ket{2}$,
for various amplitudes $V_0$ of the time-dependent absorbing potential: (I) $V_0 T=0$ (II) $V_0 T=10$ (III) $V_0 T=40$.}
\label{fig2}
\end{figure}

Fig. \ref{fig2} shows the same quantities
for the initial condition $\ket{\Psi(0)} = c_1 \ket{1} + c_2 \ket{2}$, $c_1=\sqrt{0.75}$ and $c_2=\sqrt{0.25}$. We have used the absorbing potential
given by Eq.(\ref{potgen}). The previous comments about the efficiency of the method remain valid.
Fig. \ref{fig2} illustrates clearly the efficiency of the chosen matrix in reproducing
the boundary conditions.

\begin{figure}[!ht]
 \includegraphics[width=\linewidth]{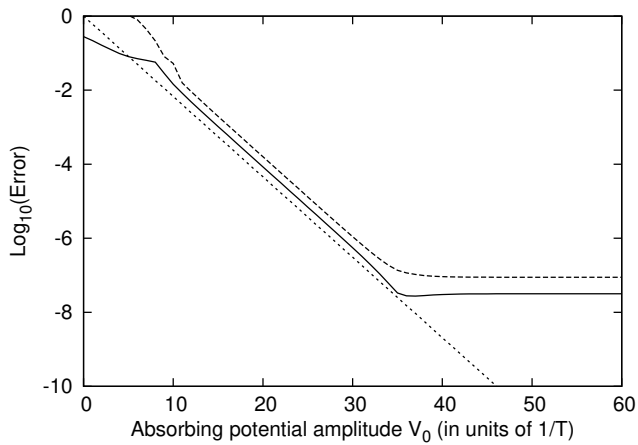}
\caption{Integrated logarithmic error estimation between direct integration and CATM with 512 time grid points as a function of $V_0$,
calculated with the first component $\langle 1 \ket{\Psi}$ for the 2-level system (i).
Errors on population, see Eq. \eqref{epsp} (solid line), and on angles, see Eq. \eqref{epsa} (dashed line). We remark that these errors follow the anticipated exponential law  Eq. \eqref{acc2} (dotted line) until they reach a plateau due to grid effects of CATM}
\label{fig6}
\end{figure}

We now give a more precise analysis of how the exact solution is approached.
To this end we define a measure of the difference between the CATM results and the direct integration results.
For the single component $\langle 1 \ket{\Psi}$ calculated by the two methods we define the integrated difference of population and
of angle:
\begin{subequations}
\begin{eqnarray}
\label{epsp}
 \epsilon_{p} &=& \frac{1}{T}\int_0^T \left(\vert \langle 1
 \ket{\Psi(t)}_{\text{CATM}}  \vert ^2 - \vert \langle 1 \ket{\Psi(t)} \vert ^2\right) dt \\
\label{epsa} \epsilon_{a}& =& \frac{1}{T}\int_0^T \left[ \arg\left( \langle 1 \ket{\Psi(t)}_{\text{CATM}} \right) -
\arg \left( \langle 1 \ket{\Psi(t)} \right) \right] dt\qquad\
\end{eqnarray}
\end{subequations}
These quantities are represented on Fig. \ref{fig6} as a function of the absorbing potential amplitude $V_0$.
With the logarithmic scale, we observe a quasi-linear law for $V_0\in[10,35]$ in consistency with Eq. \eqref{acc2}.
The error estimates next reach plateaus which are interpreted by the grid effects due to the finite basis representation of the time in the CATM method.
Indeed we can lower the level of the plateaus by increasing the number of the grid points (not shown).
%The Fig was obtained with 512 time grid points and 8 times more time steps for the direct propagation.
%It appears that the halting of the decrease of estimated error happens when
%the CATM and direct propagation errors become of the same order of magnitude and leads to a plateau which fails to represent correctly
%the way in which the CATM results and the exact ones differ by an exponentially decreasing function of the increasing amplitude of the absorbing %potential.

\subsubsection{Three-state model}

\begin{figure}[!ht]
 \includegraphics[width=\linewidth]{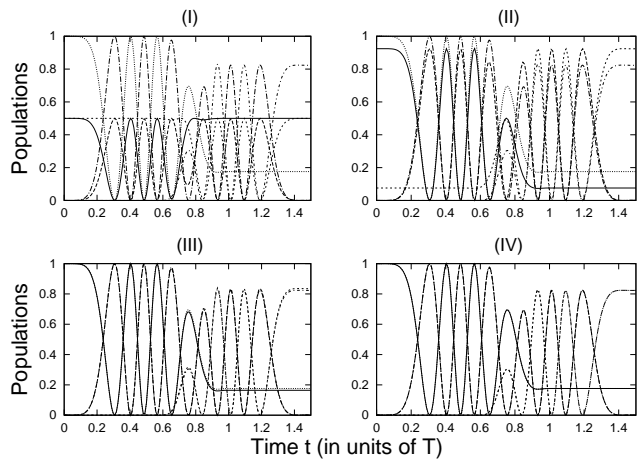}
\caption{Evolution of the populations $p_n(t)$ the three-level model (iii); exact (numerical) results $p_1$ (dots), $p_2$ (long dot-dashes), $p_3$ (dot-dashes)
and CATM results $p_1$ (solid line), $p_2$ (long dashes), $p_3$ (short dashes) without detuning and for various amplitude of the absorbing potential,
I: $V_0 T_1=0$, II: $5$, III: $10$, IV: $40$. }
\label{fig3}
\end{figure}

\begin{figure}[!ht]
 \includegraphics[width=\linewidth]{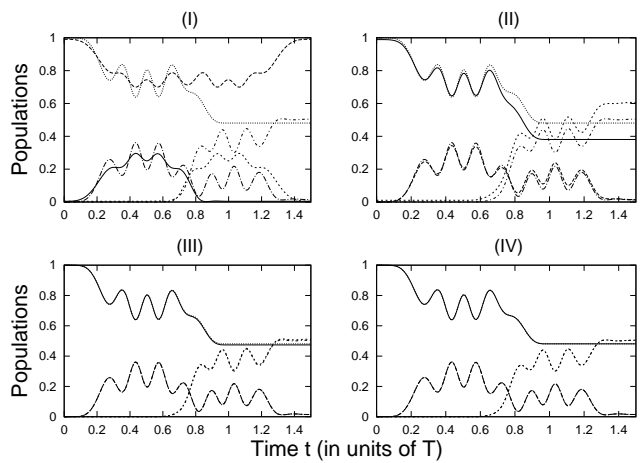}
\caption{Same as Fig. \ref{fig3}, but with a detuning $\Delta_0T_1=20$.}
\label{fig4}
\end{figure}

For the 3 level system the evolution of the population in the three-level model (iii) (as defined in section \ref{exemples})
is shown in Fig. \ref{fig3} and Fig. \ref{fig4}, without or with detuning ($\Delta_0 T_1=0$ or $\Delta_0T_1 =20$).
The selected field amplitude was $\Omega_0 T_1 = 20$ and the absorbing potential was gradually turned on from $V_0T_1=0$ to $V_0T_1=40$.
Here again, if the absorption is not sufficient, the results are wrong.
%The last frame with $V_0=40$ can be analyzed.
%Without detuning (Fig \ref{fig3}), during the first pulse (coupling
%between levels $1$ and $2$ for $t\in[0,1]$), we observe fast inversions of populations between these
%two levels. From the beginning of the second pulse (coupling levels $2$ and $3$ for $t\in[0.5,1.5]$), $p_3$ begins to grow,
%then $p_1$ stabilizes at $\simeq0.18$, while fast inversions continue to occur but this time between levels $2$ and $3$.
%At the end $p_2\simeq 0$ while $p_3\simeq0.82$.
%With detuning (Fig \ref{fig4}), fast inversions are transformed into fast but very slight oscillations, leading finally to $p_1\simeq0.48$,
%$p_2\simeq 0.02$ and $p_3\simeq0.5$. The population inversion is less important.

\begin{figure}[!ht]
 \includegraphics[width=\linewidth]{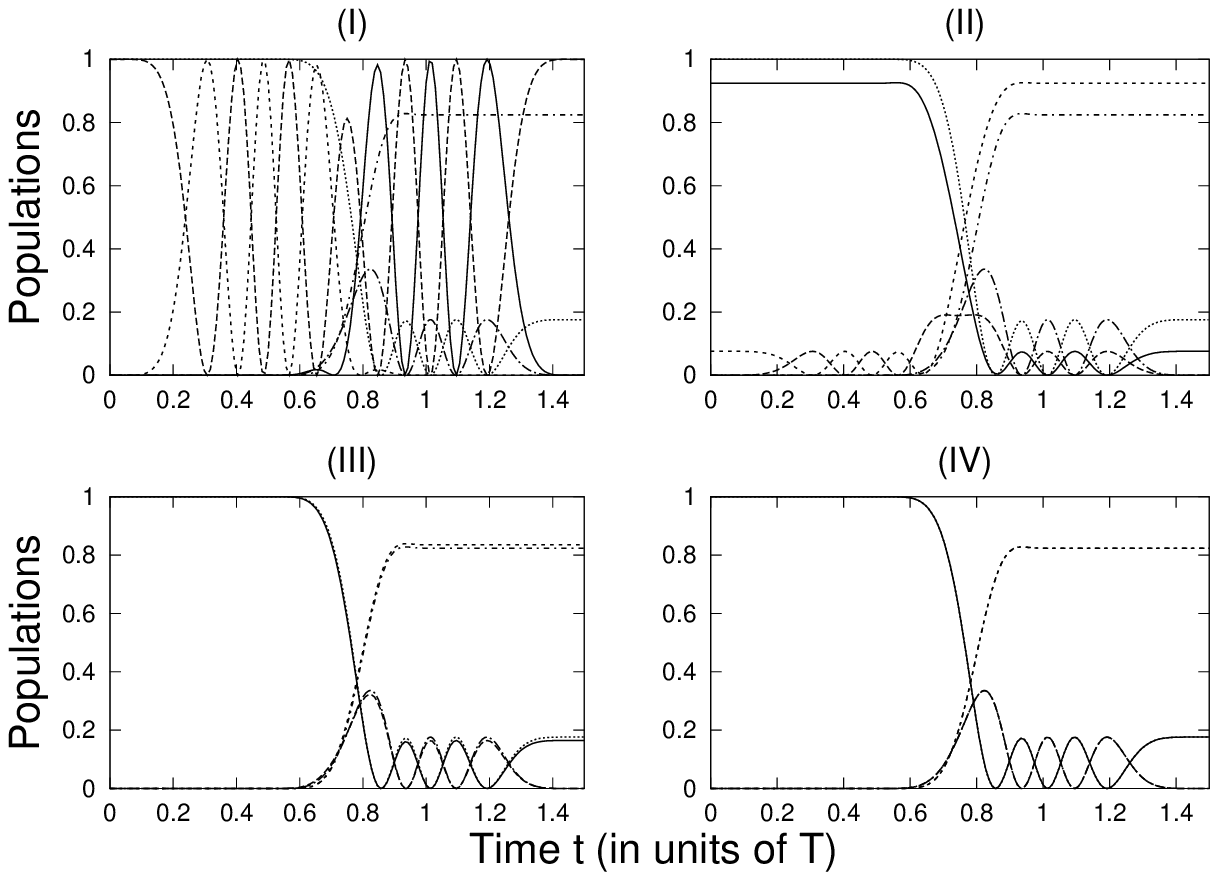}
\caption{Evolution of the populations $p_n(t)=\vert \langle n \ket{\Psi (t)} \vert ^2$ in the STIRAP model (iv) ; exact (numerical) results $p_1$ (dots), $p_2$ (long dot-dashes), $p_3$ (short dot-dashes)
and CATM results $p_1$ (solid line), $p_2$ (long dashes), $p_3$ (short dashes) without detuning and for various amplitude of the absorbing potential,
I : $V_0 T_1=0$, II : $5$, III : $10$, IV : $40$. }
\label{fig5}
\end{figure}

The results for the STIRAP model (iv) (as defined in section \ref{exemples}) are displayed in Fig \ref{fig5}.
The coupling terms between levels $2$ and $3$ are turned on before the coupling terms between $1$ and $2$ and a relatively large population inversion is observed.

\begin{figure}[!ht]
 \includegraphics[width=\linewidth]{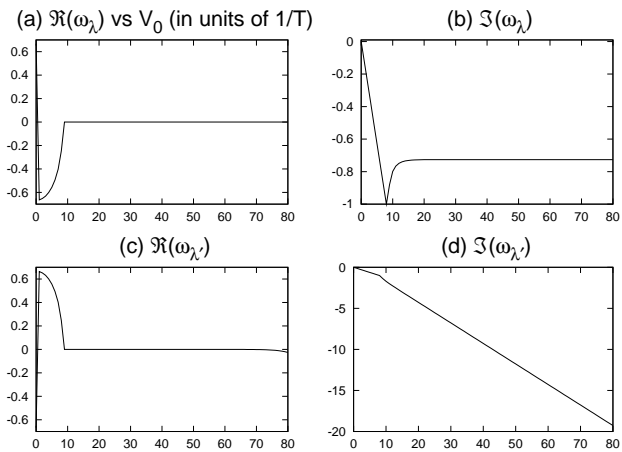}
\caption{(a) $\Re(\omega_{\lambda})$, (b) $\Im(\omega_{\lambda})$,
(c) $\Re(\omega_{\lambda'})$ and (d) $\Im(\omega_{\lambda'})$, versus $V_0$ (all in units of $1/T_1$) for the two-level model (i). When $V_0$ grows, $\omega_{\lambda'}$ moves away from $\omega_{\lambda}$ acquiring a imaginary part proportional to $V_0$.}
\label{eig1}
\end{figure}

\subsection{The expansion of the spectrum}

We now analyse the effect of dilatation of the eigenvalues by the the absorbing potential, that is the feature of separating the ``connected'' eigenvalue with respect to the other ones.
Fig. \ref{eig1} shows the Floquet eigenvalues $\{\omega_{\lambda}\}$ and $\{\omega_{\lambda'}\}$
in the first Brillouin zone calculated for the two-level model (i) as functions of $V_0$ .
Apart for small absorbing potential amplitudes where one notices an ambiguity concerning
the labelling of eigenvalue \cite{viennot}, $\Im(\omega_{\lambda})$ is a constant value in agreement with Eq. \eqref{Elambda},
and $\Im(\omega_{\lambda'})$ shows a linear evolution as predicted by
Eq. \eqref{Elambdap}.

%With our choice of optical potential, if there were no coupling terms then the theoretical rate of decrease
%$-\frac{1}{T'} \int_{T}^{T'} V_{\text{opt}} (t) dt$ would be $-1/4$, which is precisely the observed slope when %$V_0$ is large compared to the coupling parameters.

\begin{figure}[!ht]
\includegraphics[width=\linewidth]{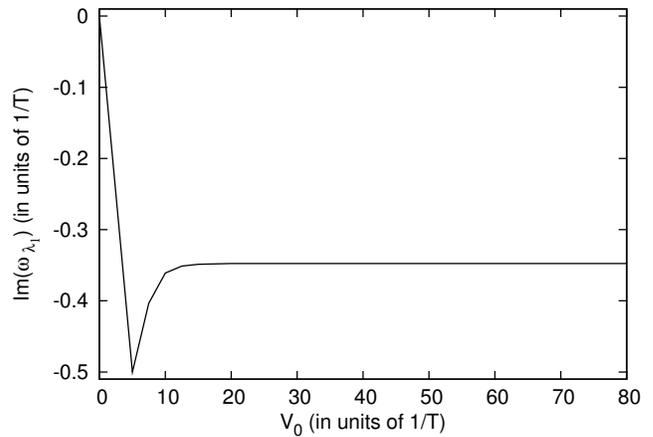}
\caption{$\Im(\omega_{\lambda_1})$  versus $V_0$ for the three-level model (iii). }
\label{eig2}
\end{figure}

\begin{figure}[!ht]
 \includegraphics[width=\linewidth]{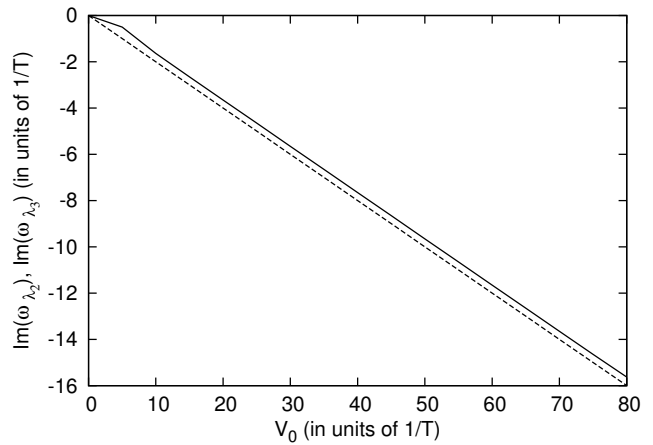}
\caption{$\Im(\omega_{\lambda_2})$ (solid line) and $\Im(\omega_{\lambda_3})$ (dashed line) versus $V_0$ for the three-level model (iii).}
\label{eig3}
\end{figure}

Figures \ref{eig2} and \ref{eig3} refer to the three-level system (iii) and show
the same features.
Concentrating on the imaginary part of the ``connected'' Floquet eigenvalue (Fig. \ref{eig2})
we see that, after a region of stabilization, $\Im(\omega_{\lambda_1})$ is no longer affected by the growth of the absorbing potential.
By contrast both $\Im(\omega_{\lambda_2})$ and $\Im(\omega_{\lambda_3})$ acquire imaginary parts which are linear with respect to $V_0$.

This feature will be useful in practice for large systems, in particular
if a wave operator method is used to find the Floquet eigenstate \cite{reviewgeorges2},
since that method is efficient in finding isolated eigenvalues.

\subsection{The influence of the number of Fourier basis functions}

\begin{figure}[!ht]
 \includegraphics[width=\linewidth]{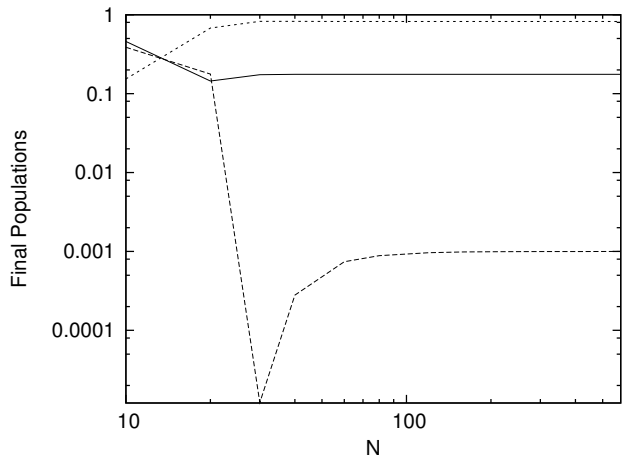}
\caption{Stability of final populations $p_1$ (solid line), $p_2$ (dashed line) and $p_3$ (dotted line) functions of the Fourier time grid point number $N$,
in logarithmic scale.}
\label{stability}
\end{figure}

We give here some details about the stability of the results
as the number of Fourier basis functions $N$ is reduced in the CATM calculation.
To increase calculational speed and to decrease memory requirement it appears necessary
to use as small value of $N$ as possible.
Figure~\ref{stability} shows how the final populations obtained in the CATM calculation
varies as $N$ is increased. These calculations correspond to the STIRAP model (iv).

The values $p_3 \simeq 0.82$ and $p_1 \simeq 0.18$ are stable for $N \gtrsim 30$ but $p_2\simeq 10^{-3}$ is not obtained
accurately until $N$ is about $80$. We see that finding small  probabilities in absolute values requires a more precise description of the temporal evolution;
however about 80 grid points appear to be ample for the calculations. The general principle is to choose an $N$ which is high enough to
follow the time variations in the Hamiltonian and to obtain accurate values for small probabilities.

\section{conclusion}

The optimum computational implementation of the CATM is still being actively researched but the basic principles behind the method are simple to follow. A static absorbing potential is often used in treating the time development of a wavefunction within the Floquet formalism. The novelty of our approach is that the absorbing potential is given a time-dependent form such that it actively constrains the wavefunction, both by imposing the correct boundary conditions on it and by modifying the spectrum so that the specific eigenvalue which is appropriate to describe the dynamical process is rendered relatively isolated from the other eigenvalues. The dynamical problem is then rendered into an eigenvalue problem in which the isolated eigenvalue is easier to find by techniques such as the Bloch wave operator method. That it is indeed possible to choose the time-dependent potential so as to produce the favourable features described above has been demonstrated for two small-scale systems for which accurate comparison results are available. For these small test systems the CATM gives accurate results, although it is clear that the eigenvalue problems which arise can involve strongly non-Hermitian matrices.

The CATM has some formal advantages for systems with a time-dependent Hamiltonian. A common approach for such systems is to use a step-by step propagation procedure with very small time steps. Many time steps are thus required to cover a given time interval  and this leads to an accumulation of errors as the propagation proceeds. By contrast, in the CATM the solution is global over the full time interval and so there is no accumulation of errors; this feature is similar to that shown by the $(t,t')$ method \cite{peskin}. 
As expected (and confirmed by the present study) the accuracy achievable within the CATM is governed by the ability to reproduce the initial conditions by suitably adjusting the time-dependent potential and by the use of a sufficiently dense Fourier time grid to describe any fast time variations contained in the Hamiltonian.

Our model calculations have also made clear the role of the time-dependent absorbing potential in dilating the Floquet spectrum so that the dynamical problem of propagation within a Hilbert space of a given dimension can be converted to that of locating an isolated eigenvalue of a  non-Hermitian matrix of much larger dimension.
The difficulty of solving the dynamical problem is thus converted into the technical problem of devising efficient algorithms for large non-Hermitian matrices. In Ref. \cite{CATM}
a previous version of the CATM was successfully tested on a molecular system involving a few hundred states. At the moment we believe that the task of isolating and then calculating the important dynamically relevant complex eigenvalue is probably not possible within the CATM for systems which are much larger than those treated in \cite{CATM}; nevertheless, the method may be useful for some systems which cause difficulties for the usual propagation methods.

\begin{acknowledgments}
We acknowledge the support from the French Agence Nationale de la
Recherche (Project CoMoC), the European Marie Curie Initial Training
Network Grant No. CA-ITN-214962-FASTQUAST, and from the Conseil
R\'egional de Bourgogne.
\end{acknowledgments}

\appendix
\section{Properties of dissipative propagators \label{appA}}

We consider a traceless time-dependent dissipative Hamiltonian $H_T$, i.e. having complex diagonal elements (with negative imaginary parts) and being self-adjoint when it is restricted to its non-diagonal elements. It has the corresponding propagator $U_{H_T}(t,t_0)$: $i\frac{\partial}{\partial t}U_{H_T}(t,t_0)=H_T U_{H_T}(t,t_0)$, and its adjoint $H_T^{\dagger}$ the propagator $U_{H_T^{\dagger}}(t,t_0)$. They satisfy
\begin{equation}
\det[U_{H_T}(t,t_0)]=\det[U_{H_T^{\dagger}}(t,t_0)]=1.
\end{equation}
>From the definition of the propagators, we obtain $\frac{\partial}{\partial t}[U^{\dagger}_{H_T^{\dagger}}(t,t_0)U_{H_T}(t,t_0)]=0$, that is
\begin{equation}
U^{\dagger}_{H_T^{\dagger}}(t,t_0)U_{H_T}(t,t_0)=\openone.
\end{equation}
For the two-state case of general Hamiltonian
\begin{equation}
\label{H} H=\left(\begin{array}{cc} \Delta_1 & \Omega \\ \Omega^{\ast} & \Delta_2\end{array}\right)
\end{equation}
with in general complex $\Delta_1$ and $\Delta_2$ and time-dependent parameters: $\Delta_j\equiv\Delta_j(t)$, $\Omega\equiv\Omega(t)$, we first decompose it as a term proportional to identity and a traceless term:
\begin{equation}
H=\frac{\Delta_1+\Delta_2}{2}\openone
+H_T.
\end{equation}
with
\begin{equation}
H_T=\left(\begin{array}{cc} -\frac{\Delta_2-\Delta_1}{2} & \Omega \\
\Omega^{\ast} & \frac{\Delta_2-\Delta_1}{2}\end{array}\right).
\end{equation}
The propagator for $H$ reads
\begin{equation}
U_H(t,t_0)=e^{-i\frac{1}{2}\int_{t_0}^{t}dt'(\Delta_1(t')+\Delta_2(t'))}
U_{H_T}(t,t_0).
\end{equation}
If $\Delta_1$ and $\Delta_2$ are real (non dissipative self-adjoint Hamiltonian), the propagator $U_{H_T}(t,t_0)$ is unitary and is thus of the form:
\begin{equation}
U_{H_T}(t,t_0)=\left(\begin{array}{cc} a & -b^{\ast} \\ b & a^{\ast}\end{array}\right).
\end{equation}
In the general case of complex $\Delta_1$ and $\Delta_2$, this is not true anymore. We write the propagator as
\begin{equation}
U_{H_T}(t,t_0)=\left(\begin{array}{cc} a & c \\ b & d\end{array}\right),\quad ad-bc=1.
\end{equation}
For the adjoint of $H$:
\begin{equation}
H^{\dagger}=\frac{\Delta_1^{\ast}+\Delta_2^{\ast}}{2}\openone
+H_T^{\dagger},
\end{equation}
the propagator writes
\begin{equation}
U_{H^{\dagger}}(t,t_0)=e^{-i\frac{1}{2}\int_0^{t}dt'
(\Delta^{\ast}_1(t')+\Delta^{\ast}_2(t'))}U_{H_T^{\dagger}}(t,t_0),
\end{equation}
where $U_{H_T^{\dagger}}(t,t_0)$ connects with $U_{H_T}(t,t_0)$ as
\begin{equation}
U_{H_T^{\dagger}}(t,t_0)=\left(\begin{array}{cc} d^{\ast} & -b^{\ast} \\ -c^{\ast} & a^{\ast}\end{array}\right).
\end{equation}
These properties are used to obtain a link between two wavefunctions resulting
from the two orthogonal initial states $\left(\begin{array}{ll}
1\\
 0
 \end{array}\right)$ and
$\left(\begin{array}{ll}
0\\
 1
 \end{array}\right)$
and respectively driven by $H$ and $H^{\dagger}$, i.e. $U_{H_T}(t,t_0)\left(\begin{array}{ll}
1\\
 0
 \end{array}\right)=\left(\begin{array}{ll}
a\\
 b
 \end{array}\right)$
 and $U_{H_T^{\dagger}}(t,t_0)\left(\begin{array}{ll}
0\\
 1
 \end{array}\right)=\left(\begin{array}{ll}
-b^{\ast}\\
 a^{\ast}
 \end{array}\right)$.
%\begin{subequations}
%\begin{eqnarray}
%U_H (t,0) \left(\begin{array}{ll}
%1\\
% 0
% \end{array}\right) &=& \left(\begin{array}{ll}
%a\\
% b
% \end{array}\right) e^{-i\frac{1}{2}\int_0^t dt' (\Delta_1+\Delta_2)} \\
%U_{H^{\dagger}} (t,0) \left(\begin{array}{ll}
%0\\
% 1
% \end{array}\right) &=& \left(\begin{array}{ll}
%-b^{\star}\\
% a^{\star}
% \end{array}\right) e^{-i\frac{1}{2}\int_0^t dt' (\Delta_1^{\star}+\Delta_2^{\star})}
%\end{eqnarray}
%\end{subequations}

\section{Construction of the Floquet Hamiltonian \label{appB}}
In this Appendix, we give some details about the structure of the Floquet Hamiltonian.
For the time dimension, we work with a discrete variable representation (DVR) $\{ \vert t_i \rangle \}$, $i=1\cdots N$, associated with a Fourier finite basis representation (FBR).
We show that the time derivative operator can be expressed simply in the DVR basis:
Let $I_j$ be a column vector of component $\delta_{ij}$, $i=1\cdots N$, then
\begin{equation}
 \Bigl\langle t_i \Bigr\vert -i \frac{\partial }{\partial t} \Bigl|t_j\Bigr\rangle =
 FFT^{-1}_i \left( \begin{array}{c}\omega_1 FFT_1 (I_j)\\
 \vdots\\
\omega_N FFT_N (I_j)\end{array}  \right),
\end{equation}
where $FFT_i$ represents the $i^{th}$ fast Fourier transform component and $\omega_i$
is the Fourier angular frequency defined by
\begin{equation}
\label{Fourierf}
 \left \lbrace
\begin{array}{lll}
\omega_n  =   \frac{2\pi }{T'} (n-1) \quad  1 \le n \le \frac{N}{2}, \\
\omega_n =  \frac{2\pi }{T'} (n-1-N) \quad  \frac{N}{2} < n \le N.
\end{array} \right.
\end{equation}
Due to the periodicity, \eqref{Fourierf} is equivalent to
\begin{equation}
\label{Fourierf_}
\begin{array}{lll}
\omega_n  =   \frac{2\pi }{T'} (n-1) \quad  1 \le n \le N. \\
\end{array}
\end{equation}
The matrix representation of $-i \frac{\partial }{\partial t}$  is diagonal in the molecular basis, and
$H(t)$ and ${\cal{V}}(t)$ are approximately diagonal in the DVR time basis. Consequently the Floquet Hamiltonian
for the two-level models (i) or (ii) with the initial condition $c_1=1$, $c_2=0$ is approximately represented in the $\{ \ket{1},\ket{2} \} \otimes \{ \ket{t_i} \}$ basis by :
\begin{equation}
% use packages: array
\left(
\begin{array}{lllll}
\partial_{t_{11}} & \Omega_1 & \partial_{t_{12}} & 0 &  \ldots \\
\Omega_1 & (\partial_{t_{11}}+\Delta_1-iV_1) & 0 & \partial_{t_{12}} &  \\
\partial_{t_{21}} & 0 & \partial_{t_{22}} & \Omega_2 &  \\
0 & \partial_{t_{21}} & \Omega_2 & (\partial_{t_{22}}+\Delta_2-iV_2) & \\
\vdots & & & &  \ddots
\end{array}
\right)
\end{equation}
with
\begin{eqnarray}
 \partial_{t_{ij}} &\equiv& \Bigl\langle t_i \Bigr\vert -i \hbar\frac{\partial }{\partial t} \Bigl|t_j\Bigr\rangle  \nonumber\\
\Omega_i &\equiv & \Omega (t_i) \quad \forall t_i\in [0,T] \quad(0 \text{ elsewhere})\nonumber\\
\Delta_i &\equiv & \Delta (t_i) \quad \forall t_i \in [0,T] \quad(0 \text{ elsewhere})\nonumber\\
-iV_i &\equiv & -i V_{\text{opt}} (t_i) = -i V_0 \sin ^2 \left( \frac{t_i-T}{T'-T}\right) \nonumber\\
&&\forall t_i \in [T,T'] \quad(0 \text{ elsewhere}) \nonumber
\end{eqnarray}
This construction can be directly generalized to treat three-level or larger systems.

%merlin.mbs 2010-03-15 4.21a (PWD, AO, DPC)
%Control: key (0)
%Control: author (8) initials jnrlst
%Control: editor formatted (1) identically to author
%Control: production of article title (-1) disabled
%Control: page (0) single
%Control: year (1) truncated
%Control: production of eprint (0) enabled
%

%\nocite{*}
%\begin{references}

%\end{references}

\bibliography{CATM_ss7}

\end{document}